УДК 004.9:66.013.512

# МОДЕЛИРОВАНИЕ РАЗВЕРНУТЫХ СХЕМ АВТОМАТИЗАЦИИ ТЕХНОЛОГИЧЕСКИХ ПРОЦЕССОВ В САПР РЕКОНСТРУКЦИИ ПРЕДПРИЯТИЙ

В.В. Мигунов

**Введение**

Согласно требованиям системы проектной документации для строительства (СПДС) схемы автоматизации необходимы практически в каждом проекте реконструкции промышленных зданий и сооружений, в которых осуществляются какие-либо технологические процессы. Содержание и оформление схем автоматизации в виде чертежей определяются в основном стандартами [1, 2] и руководящим материалом [3]. Наличие этих стандартов порождает возможность значительной автоматизации труда проектировщика в рамках соответствующих специализированных систем автоматизированного проектирования (САПР), независимо от того, разрабатываются ли они на базе внешних графических ядер (как приложения к AutoCAD$^{TM}$, например), или в рамках независимой САПР. Однако в литературе не встречаются сведения об устройстве компьютерных моделей чертежей схем автоматизации СПДС для целей САПР, несмотря на наличие самих САПР такого назначения.

В настоящей работе излагаются модельные представления схем автоматизации, отвечающие принципу "исключить повторный ввод информации", реализованные в САПР TechnoCAD GlassX [4] и проверенные временем при создании нескольких сотен схем. Рассмотрение ограничивается схемами, выполненными развернутым способом, в отличие от упрощенного способа [1], хотя имеется значительная общность этих видов схем, и получаемые практические результаты могут применяться для разработки также и упрощенных схем.

**1. Чертеж схемы автоматизации технологического процесса по СПДС**

В чертеже развернутой схемы автоматизации присутствуют основные надписи, тексты технических требований и другие элементы, присущие всем чертежам и не специфичные для именно этих схем. Автоматизация их создания здесь не рассматривается.

Для целей САПР выделяются три основные функционально различные существенные части схемы, показанные на рис.1 сносками от крупных контурных номеров слева.

1 - условные графические обозначения приборов и исполнительных механизмов непосредственно на технологической схеме производственного процесса с линиями связи их с контурами управления. Технологическая схема первична, она разрабатывается до подготовки схемы автоматизации и дорабатывается параллельно с этой подготовкой. Существенно, что расположение и назначение средств автоматизации (СА) в части 1 обусловлено расположением технологических устройств, с которыми они связываются;

2 - таблица расположения средств автоматизации (по месту, на щите,…), включающая условные обозначения СА и линии их связей в контурах управления. Это отдельная часть чертежа, специфичная только для СА;

3 - экспликация КИПиА, или спецификация, или заказная спецификация - один или более табличных конструкторских документов, специфицирующих СА данной схемы. Каждая строка таблицы связана с какими-либо СА в схеме через их обозначения,

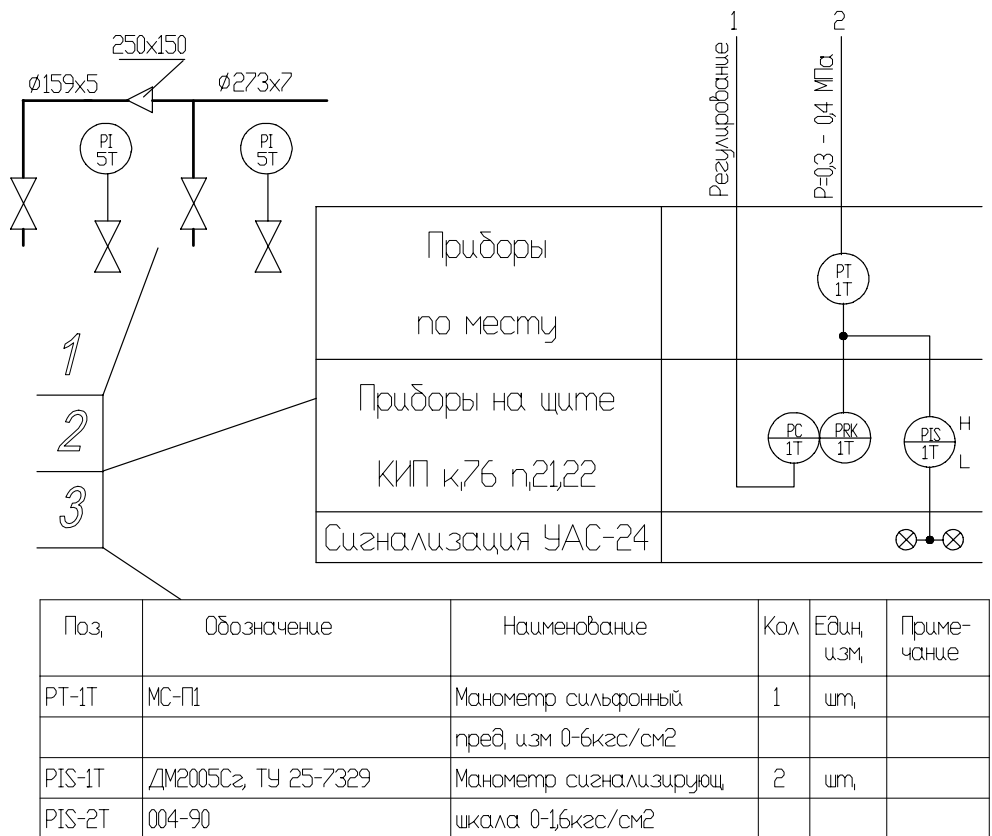

Рис.1. Фрагменты трех основных частей схем автоматизации указываемые в графе "Поз.". Специфицирующие сведения выбираются из номенклатурных каталогов выпускаемых СА или паспортов изделий. В таблице они могут группироваться в разделы, например, по контурам управления.

Согласно [2] буквенное обозначение СА жестко связано с его функциями, а графическое обозначение в виде круга или овала и дополнительных элементов имеет стандартизированные размеры. Также в [2] зафиксированы правила нанесения линий связи с соединением и без и др.

## 2. Представление элементов схем как модулей в чертеже

В силу интенсивных связей части 1 схемы с независимой от нее технологической схемой параметрическое представление не может полностью определять геометрию всех частей схемы, и применение в полном объеме модульной технологии разработки расширений САПР [5] с помещением всей схемы автоматизации в один модуль в данном случае нецелесообразно. Используются отдельные модули в чертеже различных типов. Модуль имеет дуальную природу, он включает геометрическую видимую часть и набор свойств - комплект параметров. Первичны параметры, по которым генерируется видимая часть. Как элемент чертежа, он может выбираться, переноситься, удаляться и т.д. Модуль имеет тип, и каждому типу модулей соответствует допустимое множество свойств. Ниже приводятся три используемых типа модулей с перечислением допустимых свойств:

1. "Прибор": "Привязка", "Несущая геометрия", "Позиционное обозначение", "Специфицирующие свойства", "На щите", "Функциональный признак прибора", "Верхний индекс", "Нижний индекс", "Комментарий", "Тип линии приборов КИП" - для генерации изображения, выбора в электронных каталогах и специфицирования приборов. Комплект "Специфицирующих свойств" включает "Обозначение", "Наименование", "Масса", "Примечание", "Тип, марка оборудования", "Единица измерения", "Код единиц измерения", "Код завода-изготовителя", "Код оборудования, материала", "Цена", "Наименование и технич. х-ка" - полный набор сведений для генерации спецификаций любого типа. Эти сведения заполняются автоматически при выборе СА в электронных каталогах, список которых автоматически сужается в соответствии с буквенным обозначением - функциональным признаком прибора.

2. "Исполнительный механизм": "Привязка", "Несущая геометрия", "Установка исполнительных механизмов", "Комментарий", "Позиционное обозначение", "Специфицирующие свойства" - для тех же операций с исполнительными механизмами, в условных графических обозначениях которых отражаются при генерации установки наличия ручного управления и нормального положения.

3. "Таблица КИПиА": "Несущая геометрия", "Установка таблиц КИП", "Комментарий", "Текст таблиц КИП" - генерация изображения таблиц расположения СА, определение принадлежности СА разделам при генерации спецификаций. "Установка таблиц КИП" включает:

- ширину таблицы в миллиметрах;
- наименование и высоту каждого раздела таблицы;
- признак окончания таблицы на этом листе.

С помощью модулей перечисленных типов автоматизируется создание 1 и 2 частей схемы, а также подготовка специфицирующих сведений для генерации спецификаций. Сами спецификации при этом реализованы без привлечения аппарата модулей, элемент чертежа типа "Таблица" сам хранит свое параметрическое представление (без видимой части), а его изображение генерируется при выводе на экран или на печать.

### 3. Автоматизируемые в рамках модульной модели операции подготовки схем

На первом этапе в чертеж с имеющейся технологической схемой добавляются прямоугольники, изображающие щиты, пульты, агрегатированные комплексы и т.п. Они генерируются по параметрам, задаваемым в форме ввода в соответствии с требованиями [3]. Затем в чертеж помещаются новые приборы и исполнительные механизмы. На рис. 2 показаны вводимые для приборов сведения и справа от них – сгенерированное по этим сведениям условное графическое обозначение прибора в соответствии с [2].

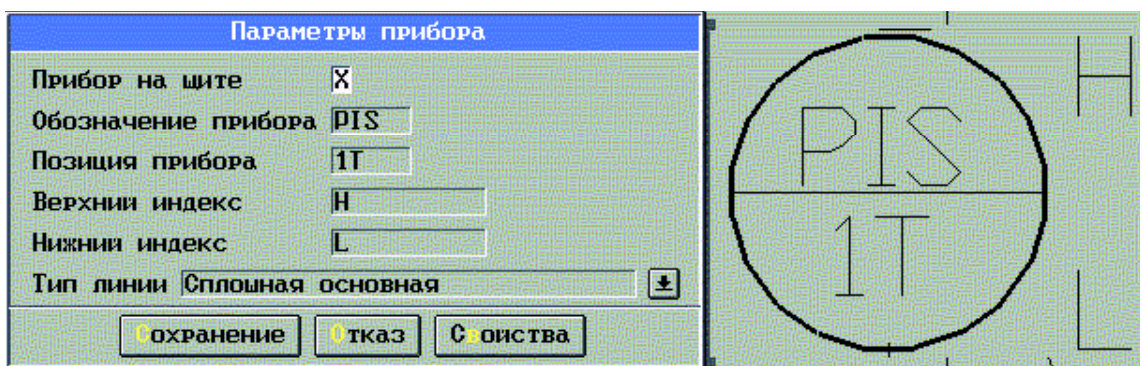

Рис. 2. Параметры прибора и его изображение

После генерации обозначения прибора предлагается провести линию связи между прибором и точкой контроля в виде ломаной с вертикальными и горизонтальными сегментами по [3]. При вычерчивании линий связи автоматически реализуется привязка к тем точкам обозначений приборов и исполнительных механизмов, из которых рекомендуется начинать линии связи по [3]; включается режим ортогонализации. Затем точки соединения линий связи изображаются и стираются автоматически в циклическом режиме указания точек пересечения линий связи (рис.3).

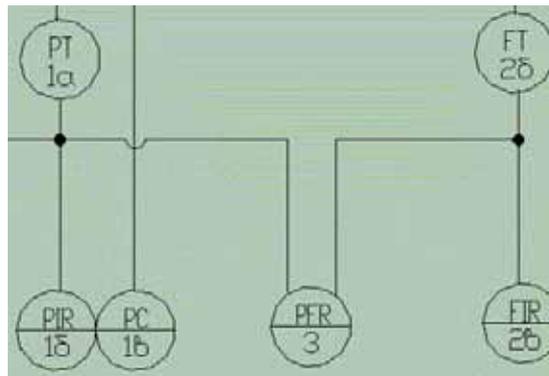

Рис. 3. Фрагмент схемы с линиями связи и точками их соединения

Как во время задания свойств нового СА, так и при изменении свойств уже имеющихся в чертеже СА, по одному и группами, их можно специфицировать путем выбора в электронных каталогах. Измеряемая величина определяется по уже имеющимся функциональным признакам, и автоматически предлагается выбор только из тех каталогов, которые содержат приборы, измеряющие эту величину. Если выбор в каталогах приводит к изменению буквенных обозначений, то соответствующие условные графические обозначения в чертеже перегенерируются.

Формирование специфицирующих табличных документов происходит автоматически, без разделов или с разделами: по месту расположения, по контурам управления. В первом случае собираются все имеющиеся в чертеже таблицы расположения, по ним определяются имеющиеся разделы и входящие в них СА. Во втором случае номер контура управления берется из числовой части позиционного обозначения. Если у прибора нет позиционного обозначения, он включается в конец таблицы, в раздел "Без номера".

**Список литературы**

1. ГОСТ 21.408-93 СПДС. Правила выполнения рабочей документации автоматизации технологических процессов. – М.: Госстандарт, 1993.

2. ГОСТ 21.404-85 СПДС. Автоматизация технологических процессов. Обозначения условные приборов и средств автоматизации в схемах. – М.: Госстандарт, 1985.

3. РМ4-2-92. Системы автоматизации технологических процессов. Схемы автоматизации. Указания по выполнению. – М.: ГПКИ "Проектмонтажавтоматика", 1992.-51с.

4. **Мигунов В.В.** TechnoCAD GlassX – отечественная САПР реконструкции предприятия. Часть 3//САПР и графика, 2004, № 6, С.34-40.

5. **Мигунов В.В.** Модульная технология разработки проблемно-ориентированных расширений САПР реконструкции предприятия / Материалы Второй международной электронной научно-технической конференции "Технологическая системотехника" (ТСТ'2003), г.Тула, 01.09.2003-30.10.2003 [Электронный ресурс] / Тульский государственный университет. – Режим доступа: http://www.tsu.tula.ru/aim/, свободный. – Загл. с экрана. – Яз. рус., англ.